\documentclass[12pt]{article}

\usepackage{tikz}
\usetikzlibrary{matrix,decorations.pathreplacing}
\usepackage{chemist}
\usepackage{setspace}
\usepackage{multirow}
\usepackage{amsmath}
\usepackage{amssymb}
\usepackage{setspace}
\usepackage{graphicx}
\date{}

\title{Stochastic Evolution of
  Coagulation-Fragmentation processes using the Accurate Chemical
  Master Equation approach}
\author{Farid Manuchehrfar$^\ast$, Wei Tian\footnote{Department
    of Bioengineering, University of Illinois at Chicago (UIC),
    Chicago, Illinois, USA.}  , Tom Chou\footnote{
    Departments of Biomathematics and Mathematics, University of
    California at Los Angeles (UCLA), Los Angeles, California}, and
  Jie Liang$^\ast$\footnote{To whom correspondence
    should be addressed, Professor. Jie Liang: jliang@uic.edu}}

\begin{document}

\maketitle

\begin{abstract}
    Coagulation and fragmentation (CF) is a fundamental process by which
particles attach to each other to form clusters while existing
clusters break up into smaller ones. It is a ubiquitous process that
plays a key role in many physical and biological phenomena. CF is
typically a stochastic process that often occurs in confined spaces
with a limited number of available particles. In this study, we use
the discrete Chemical Master Equation (dCME) to describe the CF
process. Using the newly developed Accurate Chemical Master Equation
(ACME) method, we calculate the time-dependent behavior of the CF
system. We investigate the effects of a number important factors that
influence the overall behavior of the system, including the
dimensionality, the ratio of attachment to detachment rates among
clusters, and the initial conditions.  By comparing CF in one and
three dimensions we conclude that systems in higher dimensions are
more likely to form large clusters.  We also demonstrate how the ratio
of the attachment to detachment rates affect the dynamics and the
steady-state of the system. Finally, we demonstrate the relationship
between the formation of large clusters and the initial condition.
\end{abstract}

\section{Introduction}

Coagulation and fragmentation (CF) is a fundamental process in which
particles attach to each other to form larger clusters which can also
break down into smaller ones. The general mechanism
presents itself in physical processes such as spray and aerosol
\cite{Tsantilis 2000,Goudeli 2015,Keramati 2016}, biological processes
such as filament formation and capsid protein nucleation \cite{Sept 2001, Marandi 2015}, and biomedical phenomena
such as blood clotting \cite{powers 2006,Edelstein-keshat 1998,Nurden
2018,bertsch 2017,Tarbox 2013}.

The CF problem has been the focus of numerous theoretical and
experimental studies \cite{Ziff 1980,wattis 1998,Edelstein-keshat
  1998,redner_book}. Smolukowski's equation and the mass-action based
Becker-D\"{o}ring (BD) equation have been the basis of many studies
\cite{hoze 2016,niethammer 2003,penrose 1997,wattis 1998}.  Solving
these equations usually requires an assumption of infinite system
size. However, CF often occurs in confined spaces with limited numbers
of molecules \cite{wattis 1998}. The behavior of CF in such small
systems is also intrinsically stochastic and the effects of the
discreteness in particle and cluster numbers is significant. 

In addition, the CF process lies at the heart of the blood
  clotting phenomenon \cite{Engelmann 2006}. The full
  coagulation cascade involves many molecular species and numerous
  reactions, often requiring complex models such as the ordinary
  differential equation (ODE) model of Hockin {\it et al.} (with 34
  species and 42 rates) \cite{Hockin 2002}, or an even more complex
  platlet-plasma model of \cite{Chatterjee 2010}. However, key steps involving the
  formation cluster of fibrin particles can be regarded as a CF process \cite{Guy 2005}, similar to the subject of this study.

Hockin-Mann reaction network model and classic Becker-D\"{o}ring-type
models do not incorporate discreteness
and stochasticity of the CF process, when it happens in confined space \cite{davis 2016,wattis 1998}.
However, the Chemical Master Equation (CME) approach is widely used to
address discreteness and stochasticity \cite{Gupta 2017,Sudbrack
  2015,Smadbeck 2014}. Solving the CME provides an evolving landscape
in state space while the discrete form of the CME (dCME) can account for
finite size effects \cite{cao 2013,Terebus 2014,cao 2016A}.

Monte Carlo (MC) simulation is commonly used to solve the discrete CME
\cite{D'orsogna 2012,D'orsogna 2015,kotalczyk 2017,smith
  2018}. Studies based on MC simulations can incorporate both
attachment and detachment reactions, discreteness, and stochasticity
of the processes. However, they are limited by the efficiency of
sampling and only provide trajectories obeying the dCME.  To the best
of our knowledge, there is no MC-based approach that can easily
simulate the CF across all ranges of the attachment and detachment
rates, in various dimensions, and with different initial conditions.

An alternative approach is to obtain an exact solution to the
dCME. This is made possible only by using the newly developed Accurate
Chemical Master Equation (ACME) algorithm \cite{cao 2016B}. Using
ACME, we first enumerate all the microstates reachable by the CF
process given a specific initial condition \cite{cao 2008}. We then
find the transition matrix connecting these microstates which will be
used to determine the time-evolution and steady state of the
probability distribution of the system.  Using this approach, we will
analyze how dimensionality of the system, the ratio of attachment to
detachment rates among clusters, and initial conditions affect on the CF
process.

\section{Method}
%%%%%%%%%%%%%%%%%%%%%%%%%%%%%%%%%%%%%%%%%%%%%%%%%%%%%%%%%%%%%%%%%%%%%%%%%%%%%%%%

We describe the CF process using the discrete Chemical Master Equation
(dCME) \cite{cao 2016C}. In our CF problem, there exists $N$ molecular
species $n_1, n_2, ...,n_N$ and $m$ reactions with reaction rate
constants $r_1,r_2, ..., r_m$.  The $k$-th reaction is represented as
\begin{equation}
    c_{1,k} n_1 + c_{2,k} n_2 + ... + c_{N,k} n_N \to^{r_k} 
c'_{1,k} n_1 + c'_{2,k} n_2 + ... + c'_{N,k} n_N 
    \label{eq:1}
\end{equation}
We assume the fixed-volume system is well-mixed.

The microstate of the system at time $t$ can be represented with a
vector of the copy number of each species: $x(t)=
(x_1(t),x_2(t),...,x_N(t))\in \mathbb{R}^N$.  The union of all
possible microstates of the system across all times forms the state space
of the system $S$.

The rate of the $k$-th reaction which causes the transition of the system
from microstate $j$ to microstate $i$ is defined as
\begin{equation}
    A_k(x_i,x_j)\equiv r_k \prod_{z=1}^N \begin{pmatrix}
                    x_z \\
                    c_{z,k}
                \end{pmatrix} 
    \label{eq:2}
\end{equation}
Using the definitions above, the discrete Chemical Master Equation can be written as 
\begin{equation}
    \frac{\partial p(x,t)}{\partial t} =\sum[A(x,x')p(x',t)-A(x',x)p(x,t)] 
    \label{eq:3}
\end{equation}
Here, $p(x,t)$ is the probability of the microstate $x$, and $A(x,x')$ is the 
transition rate from microstate $x'$ to microstate $x$.
We can compute the probability $p(x,t)$ from Eq.~(\ref{eq:3}) using the
Accurate Chemical Master Equation (ACME) method \cite{cao 2016B}. 

\begin{figure}
\centering
\includegraphics[width=6cm]{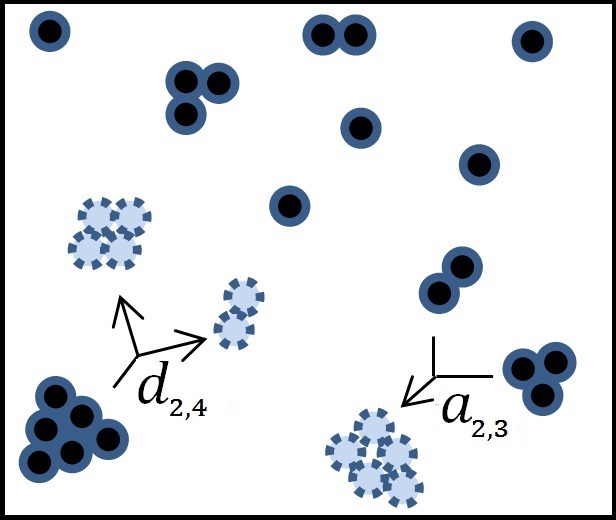}
\caption{Schematic of a CF process system showing how particles attach
  and detach.}
\label{Fig.1}
\end{figure}

In our finite-sized system, we assume there is a source reservoir of
particles with a maximum capacity of $M$.  Individual particles in the
system can be generated through a reaction that produces clusters of
size 1. Clusters of size $1$ can also be removed through a degradation
reaction, which deposits one particle back into the
source. Furthermore, a cluster of size $i$ and a cluster of size $j$
can attach to each other and form a new cluster of size ($i+j$), while
($i+j$) cannot exceed a maximum cluster size of $N$.  Clusters of
size ($i+j$) can also degrade into two clusters of size $i$ and $j$
via detachment reaction (see Fig.~\ref{Fig.1}). Thus, we have four
reactions of attachment, detachment, synthesis, and degradation
(Eq.~(\ref{eq:4}-\ref{eq:7})) in our CF system
\begin{align} \label{eq:4}
    	& X_i+X_j \reactrarrow{0pt}{0.75cm}{\ChemForm{^{a_{i,j}}}}{} X_{i+j} \ , &&  A_{att,ij}= 
    	\begin{cases} {a_{i.j}}\cdot n_i\cdot (n_j-1)/2, &	 \text{if } i=j\\ {a_{i.j}}\cdot n_i\cdot n_j,&	\text{if } i\neq j
    	\end{cases} \\ \label{eq:5}
        &X_{i+j} \reactrarrow{0pt}{0.75cm}{\ChemForm{^{{d_{i,j}}}}}{}  X_i+X_j \ , &&A_{det,ij} = {d_{i.j}} \cdot n_{i+j} \\ \label{eq:6}
        &\phi \reactrarrow{0pt}{0.75cm}{\ChemForm{^{k_s}}}{} X_1 \ ,   &&A_{s}= k_s\\ \label{eq:7}
        &X_1 \reactrarrow{0pt}{0.75cm}{\ChemForm{^{k_d}}}{} \phi \ , &&A_{d}= k_d \cdot n_1
        \end{align}
Here, $X_i$ represents a cluster of size $i$, $\phi$ the source of the
system, $n_i$ the copy number of clusters of size $i$, and
${a_{i,j}}$ and ${d_{i,j}}$ the attachment and detachment rate
constants, respectively.  For one-dimensional (1D) systems, the
clusters are linear chains of particles and the attachment and
detachment of particles occur only at the ends of the cluster.  Thus,
the attachment and detachment rates are independent of the length of
the cluster and will be taken to be constants. However, in two or
three dimensions, both the attachment and detachment rates depend on
the size of the clusters involved in the reaction.  A simple model may
be that these rates depend on the perimeter and surface area of the
clusters: ${a_{i,j}} , {d_{i,j}} \propto (i\cdot j)^{1/2}$ for 2D
systems, and ${a_{i,j}}, {d_{i,j}} \propto (i\cdot j)^{2/3}$ for
3D systems \cite{niethammer 2003}.

\begin{figure}
\centering
\includegraphics[width=8cm]{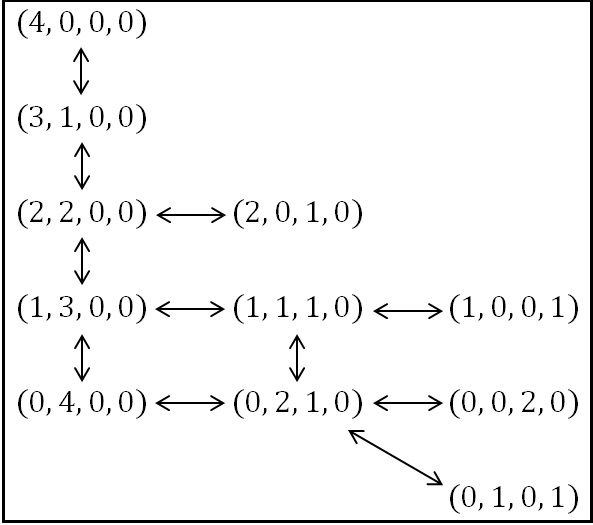}
\caption{The state space of a system with a maximum cluster size $N=3$
  and total mass $M=4$.}
\label{Fig.2}
\end{figure}

To illustrate, we give a simple example in which we have the maximum
cluster size $N=3$ and the maximum total mass of the system $M=4$. We
assume that the system starts from the initial condition where there
are 4 particles in the source and there is no cluster present in the
system. In this simple system, we can have three different types of
clusters, those of size 1, 2, and 3, respectively.
Thus, each microstate of the system can be indexed with four integers,
the first indicating the number of particles in the source; the
second, third, and forth integers indicating the number of clusters with size 1,
2, and 3, respectively.  Eq.~(\ref{eq:12}-\ref{eq:14}) are the
reactions in this simple system and the state space of the system is
illustrated in Fig.~\ref{Fig.2}.
\begin{align} \label{eq:12}
	&\phi \longleftrightarrow X_1\\ \label{eq:13}
	&2X_1 \longleftrightarrow X_2\\ \label{eq:14}
	&X_1+X_2 \longleftrightarrow X_3
\end{align}
We can then find the rate matrix and compute the probability of each
microstate (Table \ref{table.1}). From the probability of each
microstate, we can then find the expected number and the probability
of each cluster (Eq. (\ref{eq:15}) and (\ref{eq:16}), respectively)

\begin{table}
\centering

\begin{tabular}{|c|c|c|}
\hline 
State index (i) & Prob. (ACME Results) & State (Source, $n_1$, $n_2$, $n_3$) \\ \hline 
1& $p_1= 1.97 \times 10^{-1}$ & (4,0,0,0) \\ \hline  
2& $p_2= 1.97 \times 10^{-1}$ & (3,1,0,0) \\ \hline
3& $p_3= 9.84 \times 10^{-2}$ & (2,2,0,0) \\ \hline
4& $p_4= 3.28 \times 10^{-2}$ & (2,0,1,0) \\ \hline
5& $p_5= 8.20 \times 10^{-3}$ & (1,3,0,0) \\ \hline
6& $p_6= 4.92 \times 10^{-2}$ & (1,1,1,0) \\ \hline
7& $p_7= 9.84 \times 10^{-2}$ & (1,0,0,1) \\ \hline
8& $p_8= 9.84 \times 10^{-2}$ & (0,4,0,0) \\ \hline
9& $p_9= 9.84 \times 10^{-2}$ & (0,2,1,0) \\ \hline
10& $p_{10}= 9.84 \times 10^{-2}$ & (0,0,2,0) \\ \hline
11& $p_{11}= 2.46 \times 10^{-2}$ & (0,1,0,1) \\ \hline

\end{tabular}

.\onehalfspacing\caption{Probability of each micro state for our sample example}
\label{table.1}
\end{table}

\begin{align} \label{eq:15}
	&\langle n_i \rangle =\sum_l {P_l\cdot n_i}\\ \label{eq:16}
	&P_{n_i}=\sum_l P_l(n_i=0)
\end{align}
where $l$ is the microstate index (Table \ref{table.1}), $\langle n_i\rangle$ 
the expected number, and $P_{n_i}$  the probability of observing a cluster of size
$i$.

In our study, we shall restrict ourselves to a system with total mass
$M=48$ and a maximum possible cluster size $N=16$, which are much
larger than the parameters in previous studies ($M=32$, $N=8$)
\cite{D'orsogna 2015}. To describe the CF system, our state space
includes $>700,000$ microstates. For our calculations, we use a machine with a 20-core Xeon
  E5-2670 CPU of 2.5GHz, with a cache size of 20MB and 128GB Ram.
  Computing the steady state distribution at a specific ratio of the
  attachment to detachment rates (${a_{i,j}/d_{i,j}}$) takes about 38
  minutes. Computing the time-evolving probability distribution
  takes between 2,729 min and 3,292 min. Table \ref{table.2} provides details on
  the computational cost.

\begin{table}
\centering

\begin{tabular}{|c|c|c|}
\hline 
${a_{ij}/d_{ij}}$ & Steady state cost (min) & Time-evolving cost (min) \\ \hline

$3.0$& $38$ & $3,474$ \\ \hline
$4.0$& $38$ & $3,292$ \\ \hline
$5.0$& $38$ & $3,152$ \\ \hline
$10.0$& $38$ & $3,044$ \\ \hline
$20.0$& $38$ & $2,913$ \\ \hline
$30.0$& $38$ & $2,808$ \\ \hline
$40.0$& $38$ & $2,756$ \\ \hline
$50.0$& $38$ &  $2,729$\\ \hline

\end{tabular}

.\newline\caption{Computational cost for solving the steady state and the time-evolving dynamics of the system}
\label{table.2}
\end{table}

\section{Results}

Our results are organized as follows. We first examine the effect of
dimensionality on the formation of the largest cluster in the system. We
then study the effect of different attachment/detachment
rate ratios on the formation of clusters and their steady-state distributions.
Finally, we examine the
effect of different initial conditions on CF dynamics.

\subsection{Effects of Dimensionality}

\begin{figure}[!htbp]
\centering
\includegraphics[width=10cm]{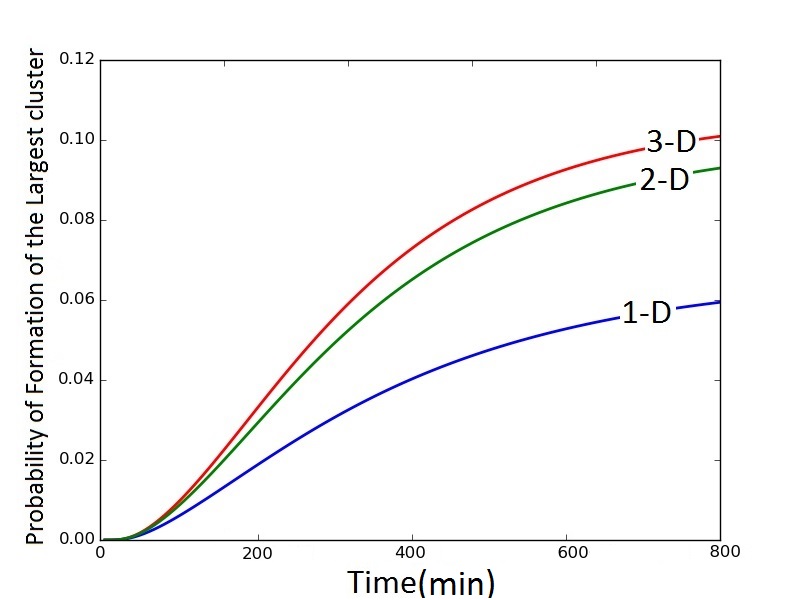}
\caption{The probability of formation of the largest cluster at different times in different dimensions when attachment/detachment rates ratio is equal to 1. 
In 2D/3D systems, this probability is twice of that in 1D while the difference between 2D and 3D systems are negligible.}
\label{Fig.3}
\end{figure}

For 1D systems, the attachment rate ${a_{i,j}}$ and the detachment
rate ${d_{i,j}}$ are independent of the size of clusters. For 2D and
3D systems, we will assume ${a_{i,j}} \& {d_{i,j}} \propto (i\cdot
j)^{1/2}$ and ${a_{i,j}} \& {d_{i,j}} \propto (i\cdot j)^{2/3}$,
respectively \cite{niethammer 2003}.  Fig.~\ref{Fig.3} compares
probability of the largest clusters at different time when
${a_{i,j}}/{d_{i,j}}=1$ in systems with different
dimensionality. There is significant difference between the 1D system
and 2D/3D systems. At long times, the probability of forming
  largest clusters in 3D is approximately twice of that in 1D. Since
the difference in the large-cluster formation probabilities is
negligible between 2D and 3D systems, we will use the 3D results for
the rest of this paper.

\subsection{Steady State Distributions}

\begin{figure}[!htbp]
\centering
\includegraphics[width=14cm]{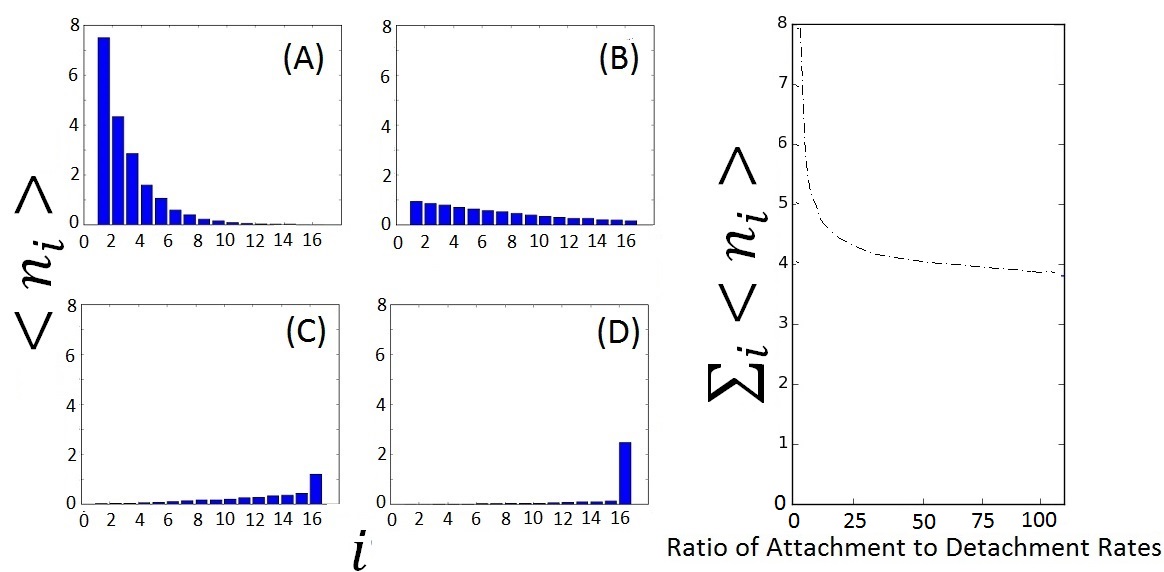}
\caption{Expected number of clusters for different
  ${a_{i,j}}/{d_{i,j}}$ at steady state, (A)
  ${a_{i,j}}/{d_{i,j}}=0.1$, (B) ${a_{i,j}}/{d_{i,j}}=1$, (C)
  ${a_{i,j}}/{d_{i,j}}=10$, (D) ${a_{i,j}}/{d_{i,j}}=1000$. 
  When ${a_{i,j}}/{d_{i,j}}$ increases, expected number of large particles in the system increases. 
(E) Expected number of clusters of all sizes in the system.}
\label{Fig.4}
\end{figure}

\textbf{Expected number of clusters}. Fig.~\ref{Fig.4}A-D shows the
expected number of clusters of different sizes for four different
values of ${a_{i,j}}/{d_{i,j}}$: $0.1$, $1$, $10$, and
$1000$. The inset shows the distribution of clusters of
different sizes at the steady state. 
When ${a_{i,j}}/{d_{i,j}}\ll 1$, all clusters are singletons. When
${a_{i,j}}/{d_{i,j}}$ increases, larger clusters form. When
${a_{i,j}}/{d_{i,j}}\approx 1000$, all clusters are at their
maximum allowed size. The expected number of all clusters at different
ratios of attachment to detachment rates is shown in
Fig.\,\ref{Fig.4}E.

\begin{figure}[!htbp]
\centering
\includegraphics[width=10cm]{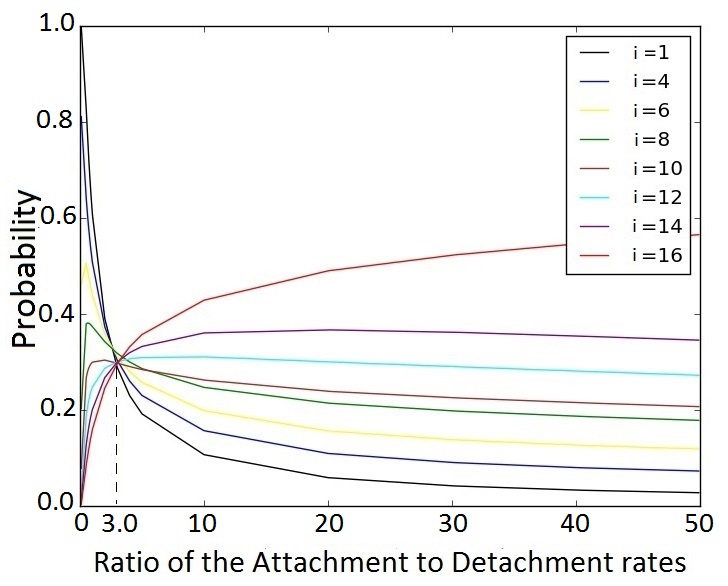}
\caption{Probability of local clusters with size $i$ at steady state for different
  ratios of the attachment to detachment rates.The probability of cluster with size 16 become more than the probability of other clusters when attachment/detachment rate ratio $>3$ while it is less than the probability of other clusters when attachment/detachment rate ration $<3$}
\label{Fig.5}
\end{figure}

\textbf{Probability of forming clusters of different sizes}.  The
formation of large clusters is an important issue in CF
processes. Without loss of generality, we set a critical
probability of having the largest cluster ($p_{16}$) to be $0.3$.
Fig.~\ref{Fig.5} shows the steady state probabilities of different clusters with
different ${a_{i,j}}/{d_{i,j}}$.  When
${a_{i,j}}/{d_{i,j}}<3.0$, $p_{16}$ is less than the probability
of other clusters ($p_{1}-p_{15}$)(Fig.\,\ref{Fig.5}A). When this
ratio is around $3$, the probabilities for all clusters are almost
equal. Thus, for the assumed threshold, ${a_{i,j}}/{d_{i,j}}=3.0$
is the critical ratio of attachment to detachment rate. Below this
value, forming the largest cluster is unlikely.

\subsection{Dynamical Behavior of the CF System}

\begin{figure}
\centering
\includegraphics[width=14cm]{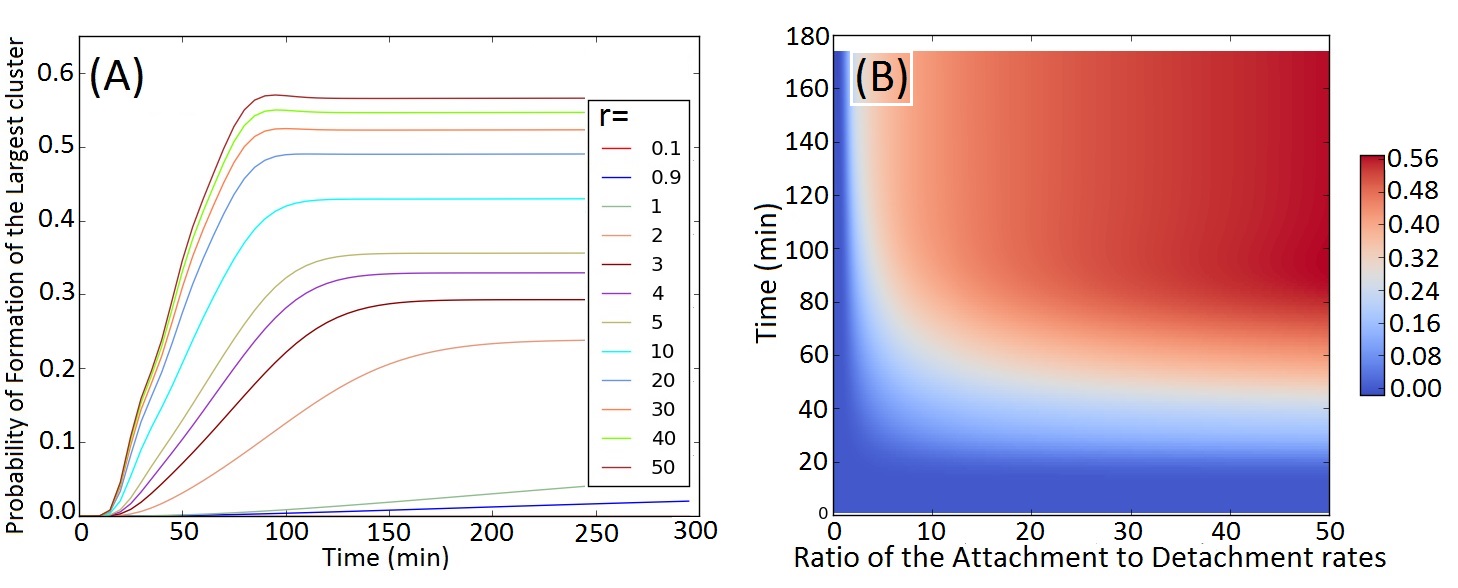}
\caption{Time-dependence of the probability to form maximum-size clusters for different ratios $r$ of the attachment to detachment
  rates. (A): Probability of formation of largest cluster grows when $r$ increases. When ${a_{i,j}}/{d_{i,j}}=3$, probability of formation of largest cluster becomes equal to the critical probability of $0.3$, after the system reaches the steady state. (B): Critical time (white region) at which the probability of forming the largest cluster reaches 0.3. Before this critical time (blue region), formation of the largest cluster is unlikely and after this critical time (red region), it is high probable that system contains largest cluster}

\label{Fig.6}
\end{figure}

The time a CF system needs to reach the critical probability of
  $p_{16}$ is a quantity of interest.  We therefore examine the
  dynamics of the system to understand the time-dependence of forming
  large clusters.  Fig.\,\ref{Fig.6}A shows how $p_{16}$ grows for
  different ratios of attachment to detachment rates. When
  ${a_{i,j}}/{d_{i,j}}<3$, the probability of forming the largest
  cluster is less than 0.3, regardless of how much time has past.
  Fig.\,\ref{Fig.6}B shows the critical time at which the probability
  of forming the largest cluster reaches 0.3 (white region). Before
  this critical time, formation of large clusters is unlikely to occur
  (blue region). A system containing large clusters are more likely
  after this critical time (red region).  In extreme cases when
${a_{i,j}}/{d_{i,j}}>1000$, it is highly probable that large
clusters will form in the system within $40$ minutes. In contrast, when ${a_{i,j}}/{d_{i,j}}
\approx 3.0$, it takes about 150 minutes for the
 system to form, with appreciable probability $>0.3$, a maximum-size cluster (Fig.\,\ref{Fig.6}B).

We examined the convergence behavior in reaching the
  steady state distribution. Table.\ 3 lists the distance to the
  steady state measured as $\arrowvert p_{16}(\infty)-p_{16}(t)\arrowvert$ 
  at different time and with different ${a_{i,j}}/{d_{i,j}}$.
  Larger ${a_{i,j}}/{d_{i,j}}$ leads to faster convergence.

\ \\
\begin{table}
\label{tf}
\centering
\begin{tabular}{|c|c|c|c|c|c|c|c|c|c|}
\hline
\multirow{2}{*}{${a_{i,j}}/{d_{i,j}}$} & \multicolumn{7}{c|}{$\arrowvert p_{16}(\infty)-p_{16}(t)\arrowvert$}                                                                                                                                                              & \multirow{2}{*}{ $p_{16}(\infty)$}  \\ [5pt] \cline{2-8}
                                        & {t=20}
                                       & {t=40}    & {t=60}   & {t=80}   & {t=100}  & {t=120}  & {t=140}  &                                 \\ [5pt ]\hline
{2.0}                & {0.244}  & {0.229} & {0.196}  & {0.160} &{0.121}  & {0.084}  & {0.050}   &{0.248}      \\ [5pt] \hline
{3.0}                & {0.292}  & {0.254}  & {0.196}  & {0.135} &{0.075}  & {0.035} & {0.012} & {0.298}      \\ [5pt] \hline
{4.0}                & {0.321} & {0.262}  & {0.185} & {0.110} & {0.047} &{0.014} & {0.000}      &{0.331}      \\ [5pt] \hline
{5.0}                &{0.343} &{0.266}  & {0.183}  & {0.095} & {0.035}  &{0.010}   &{0.000}      &{0.358}      \\ [5pt] \hline
{10.0}               &{0.402}  &{0.277}   & {0.158} & {0.056} & {0.008} &{0.000}      &{0.000}      &{0.429}      \\ [5pt] \hline
{20.0}               &{0.453}  &{0.296}  & {0.140}   & {0.034}  & {0.000}      &{0.000}      & {0.000}      &{0.492}      \\ [5pt] \hline
{30.0}               & {0.115}  &{0.306}  & {0.135}  & {0.023}  & {0.000}      &{0.000}      & {0.000}      &{0.525}      \\ [5pt] \hline
{40.0}               &{0.125}  &{0.318}  & {0.130}   & {0.019}  & {0.000}      &{0.000}      &{0.000}      &{0.550}       \\ [5pt] \hline
{50.0}               &{0.136}  &{0.327}  & {0.140}   & {0.019}  & {0.000}      &{0.000}      &{0.000}      &{0.571}      \\ [5pt] \hline
\end{tabular}

{ .\newline \caption{The convergence behavior of the system at different time steps at different ratios of the attachment to detachment rates}}

\end{table}

Following \cite{Marandi 2015}, we analyzed the elasticity and
  sensitivity of parameters of the CF system for a subset of clusters present at different
  ratio of $a_{i,j}/d_{i,j}$. Here, we use sensitivity to examine the response of the expected number of different clusters to changes in $a_{i,j}/d_{i,j}$. We use elasticity to examine the relative changes in the expected number of different clusters with respect to the relative changes in $a_{i,j}/d_{i,j}$. Following \cite{Marandi 2015}, the
  sensitivity $Se$ and the elasticity $El$ of forming a cluster of
  size $i$ are calculated as
  
\begin{equation}
    Se=\frac{\partial\langle n_i \rangle}{\partial (a_{i,j}/d_{i,j})}
    \label{eq:Se}
\end{equation}

\begin{equation}
    El=\frac{\frac{\partial\langle n_i \rangle}{\langle n_i \rangle}}{\frac{\partial(a_{i,j}/d_{i,j})}{(a_{i,j}/d_{i,j}}}=\frac{(a_{i,j}/d_{i,j})}{\langle n_i \rangle}\cdot Se
    \label{eq:El}
\end{equation}

\begin{figure}
\centering
\includegraphics[width=14cm]{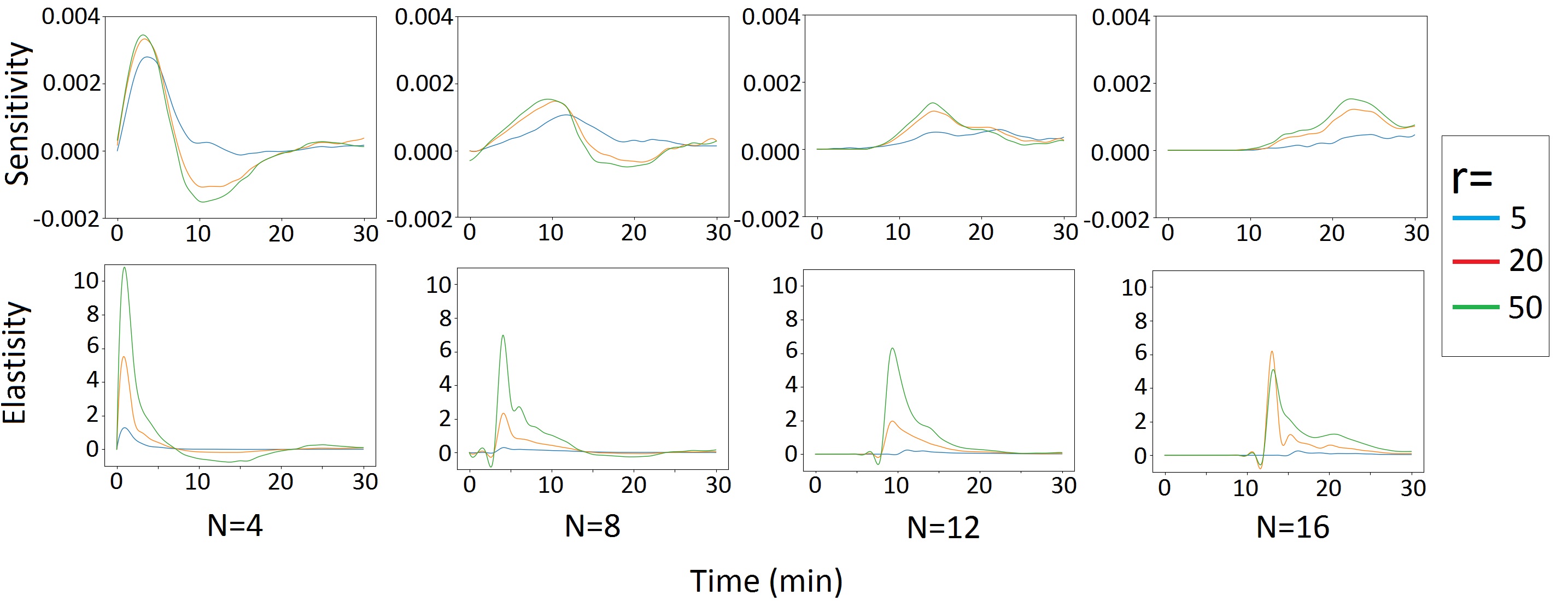}
\caption{ Sensitivity and elasticity of clusters of size $N=4,8,12,16$ in response to the ratio $r$ of the attachment to detachment rate $r=5,20,50$: when the size of cluster increase, sensitivity and elasticity of the cluster decrease.}
\label{Fig.Sen-El}
\end{figure}

Fig. \ref{Fig.Sen-El} shows $Se$ and $El$ for formation of 4 clusters of sizes $
    4, 8, 12,$ and $16$ at 3 different $a_{i,j}/d_{i,j}$ of
    $5, 20,$ and $50$.  When $a_{i,j}/d_{i,j}$ increases, $Se$ and
    $El$ decreases for forming clusters of different sizes.  In
    addition, we found that smaller clusters have higher $Se$ and
    $El$.

\subsection{Dependence on Initial Conditions}

In the examples above, we assumed 48 particles are initially in the
source which can be transported into the system through synthesis
reactions. We now examine the effect of different initial conditions
on the formation of the maximum-sized clusters and the time it takes
for the system to approach steady state.

\begin{figure}
\centering
\includegraphics[width=9cm]{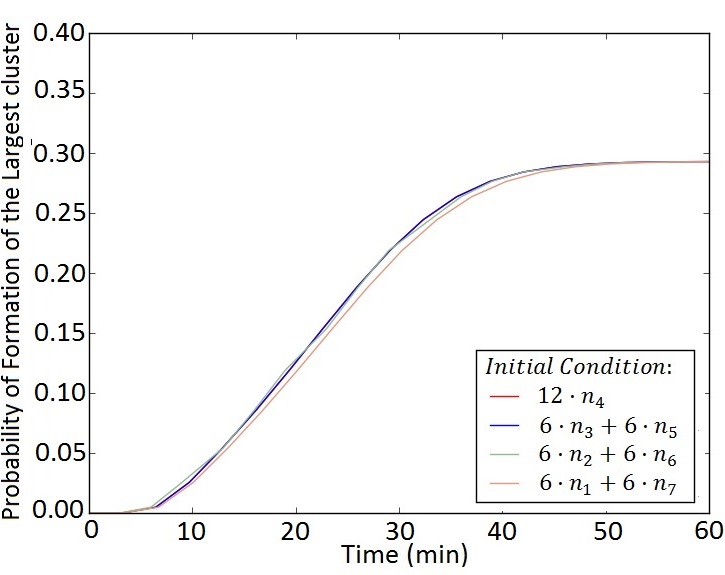}
\caption{Probability of formation of largest cluster for different
  initial conditions where initial conditions have same mean cluster size.
  When IMSs are the same, different initial conditions show very similar dynamics}
\label{Fig.7}
\end{figure}

We start with different initial conditions constrained to having the
same initial mean size (IMS) of clusters. Fig.~\ref{Fig.7} shows the
evolutions of the probability of formation of the largest cluster for
four different initial conditions: $12$ clusters of size $4$ ($12\cdot
n_4$), $6$ clusters of size $3$ and $6$ clusters of size $5$ ($6\cdot
n_3+6\cdot n_5$), $6$ clusters of size $2$ and $6$ clusters of size
$6$ ($6\cdot n_2+6\cdot n_6$), $6$ monomers and $6$ clusters of size
$7$ ($6\cdot n_1+6\cdot n_7$). All these initial conditions have the
same mean cluster size of $4$.  When the IMSs are the same, systems
with different initial conditions show very similar dynamics.
Fig.\,\ref{Fig.7} shows the behavior of the system in formation of
local clusters when the initial conditions have same mean size of
clusters.

\begin{figure}
\centering
\includegraphics[width=13cm]{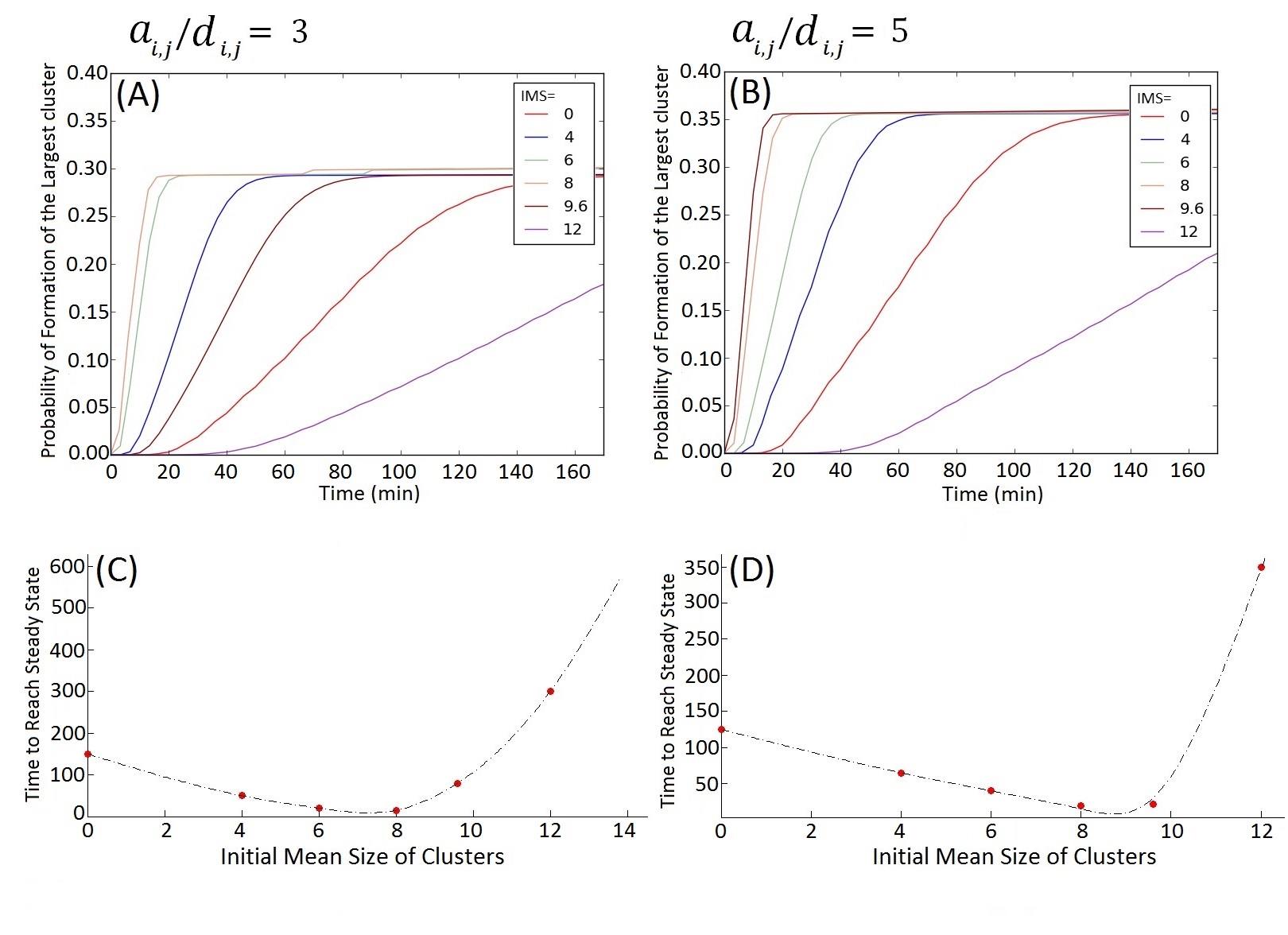}
\caption{A-B: Probability of forming maximum-sized clusters for
  different initial conditions with different initial mean size of
  clusters at ${a_{i,j}}/{d_{i,j}}=3$ and
  ${a_{i,j}}/{d_{i,j}}=5$, respectively; C-D: Time that requires
  for CF system to reach steady state at for different initial
  conditions at ${a_{i,j}}/{d_{i,j}}=3$ and
  ${a_{i,j}}/{d_{i,j}}=5$, respectively.}
\label{Fig.8}
\end{figure}

For systems with different IMSs, the dynamics on the ``distance'' of
the IMS from the steady state mean cluster size distribution (shown in
Fig.\,\ref{Fig.4}E).  Figs.\,\ref{Fig.8}A-B show the time-dependent
behavior of the probability of forming the largest cluster under
initial conditions with different IMSs and
${a_{i,j}}/{d_{i,j}}=3.0$ and ${a_{i,j}}/{d_{i,j}}=5.0$,
respectively.  When ${a_{i,j}}/{d_{i,j}}=3.0$ and $5$, the mean
size of clusters at the steady state is about $8$ and $9$,
respectively (Fig.\,\ref{Fig.4}E). Figs. \ref{Fig.8}A-B show the time
required for the system to approach steady state for different IMSs. Not surprisingly, the time to approach the steady-state distribution for values of IMS that are closer to the steady-state mean cluster sizes is less.

Figs.\,\ref{Fig.8}C-D show the time required for the system to reach
the steady state for ${a_{i,j}}/{d_{i,j}}=3.0$ and
${a_{i,j}}/{d_{i,j}}=5.0$, respectively.  Our results show that the
closer the mean size of clusters at the initial condition is to that
of the steady state, the less time it takes for the system to approach
the stationary distribution. However, we observe that, qualitatively,
systems started at IMSs greater than the mean sizes at steady state
take longer to relax than those started at IMSs smaller than that at
steady state.  

When the IMS is larger than the steady state mean size of clusters,
larger detachment rates are required for the system to reach the
steady state rapidly. However, we have $a_{i,j}/d_{i,j}>1$. Thus, we
observe asymmetry on different side of the steady state
(Fig.\,\ref{Fig.8}\,C and D). The asymmetry becomes even larger when
$a_{i,j}/d_{i,j}$ increases. As a result, the time required for
reaching the steady state increases dramatically when IMS become more
than the steady state mean size of clusters.

\section{Summary and Conclusions}

The coagulation and fragmentation is a fundamental mechanism that plays a critical role in many
physical and biological processes. Here we studied the general
properties of the CF process using the Accurate Chemical Master
Equation (ACME) method \cite{cao 2016B}, which can provide accurate
solutions to the discrete Chemical Master Equation (dCME) and can
account for the stochasticity and the discreteness of the CF process.

We examined how the dimensionality of the clusters affects its
behaviors given the same intrinsic attachment and detachment rates.
Three-dimensional systems exhibit faster dynamics compared to systems
in 1D or 2D.  The dimensionality of the clusters affects the effective
rates of attachment and detachment, which will determine the speed of
particle attachmment and detachment in a cluster.

Steady-state probability distributions of cluster sizes were also
studied under varying attachment/detachment rate ratios. For a given
critical probability of emergence of maximum-sized clusters, we are
able to determine the critical ratio between the attachment and
detachment rates.  Below this critical ratio, the large cluster of
interest is unlikely to form regardless of time. For systems with
ratio larger than the critical one, we are able to calculate the time 
required for the system to form maximum-sized cluster with high 
probability \cite{FPT_REVIEW,YVINEC}.

We further studied how different initial conditions affect the
behavior of the system and find the initial mean size of the
  clusters is one of the most important factors that govern CF
  dynamics. We find that the dynamics of systems started with
  different initial configurations with the same initial mean cluster
  sizes are similar. Further investigation shows that the dynamics
  towards steady state are controlled by the deviation of the mean
  initial cluster size from the mean cluster size at steady state.

  Future studies include analysis of various processes of self-assemblies
  of different molecular and mesoscopic-particles that occur in small
  closed systems, with supply of limited number of particles. Systems with different binding mode and binding geometry can be
  explored in details. An example is the HIV-1 viral capsid nucleation process \cite{Marandi 2015}.  In addition, critical steps of the blood-clotting
  processes involving fibrin and other molecules in the blood-clotting
  process \cite{Guy 2005} can also be studied.

\section{Acknowledgments}
JL acknowledges support from the National Institute of Health (R35GM127084, 
R01CA204962-01A1, and R21 AI126308). TC acknowledges support from the Army Research Office (W911NF-18-1-0345) and the
National Science Foundation (DMS-1516675 and DMS-1814364).


\begin{thebibliography}{10}
\bibitem{Tsantilis 2000} Tsantilis S. and Pratsinis E.  \newblock {\it
  Evolution of Primary and Aggregate Particle-Size Distributions by
  Coagulation and Sintering}.  \newblock {Aiche Journal}, {\bf
  46}(2):407-415, 2000.

\bibitem{Goudeli 2015} Goudeli Erini, Eggersdoref Maximilian L. and
  Pratsinis Sotiris E.  \newblock {\it Coagulation–Agglomeration of
    Fractal-like Particles: Structure and Self-Preserving Size
    Distribution}.  \newblock {Langmuir}, {\bf 31}(4):1320-1327, 2015.

\bibitem{Keramati 2016} Keramati Hadi, Mohammad Hassan Saidi and
  Zabetian Mohammad \newblock {\it Stabilization of the Suspension of
    Zirconia Microparticle Using the Nanoparticle Halos Mechanism:
    Zeta Potential Effect}.  \newblock {Journal of Dispersion Science
    and Technology}, {\bf 37}(1):6-13, 2016.

\bibitem{Sept 2001}
David Sept and McCammon J. A.
\newblock {\it Thermodynamics and kinetics of actin filament nucleation}.
\newblock {Biophysical Journal}, {\bf 81}(2):667-674, 2001.


\bibitem{Marandi 2015}
Farrah Sadre-Marandi, Yuewu Liu, Jiangguo Liu, Simon Tavener, Xiufen Zou
\newblock {\it Modeling HIV-1 viral capsid nucleation by dynamic systems}.
\newblock {Mathematical Biosciences}, {\bf 270}:95-105, 2015.



\bibitem{powers 2006} Powers E. T. and Powers D. L.  \newblock {\it
  The Kinetics of Nucleated Polymerizations at High Concentration:
  Amyloid Fibril Formation Near and Above the Super Critical
  Concentration}.  \newblock {Biophysical Journal}, {\bf 91}:122-132,
  2006.

\bibitem{Edelstein-keshat 1998} Edelstein-Keshet Leah and Ermentrout
  G. Bard \newblock {\it Models for the length distributions of actin
    filaments: I. Simple polymerization and fragmentation}.  \newblock
  {Bulletin of Methematical Biology}, {\bf 60}, 1998.

\bibitem{Nurden 2018} Nurden Alan T.  \newblock {\it The biology of
  the platelet with special reference to inflammation, wound healing
  and immunity}.  \newblock {Frontiers in Bioscience}, {\bf
  23}:726-751, 2018.


\bibitem{bertsch 2017}
Michiel Bertsch, Bruno Franchi, Norina Marcello, Maria Carla Tesi, Andrea Tosin
\newblock {\it Alzheimer’s Disease: a Mathematical Model for Onset and Progression, Mathematical Medicine and Biology}.
\newblock {Italian Ministary of Health}, {\bf 34}(2):193-214, 2017.

\bibitem{Tarbox 2013}
Tarbox Abigail K. and Swaroop Mamta
\newblock {\it Pulmonary embolism}.
\newblock {International Journal of Critical Illness and Injury Science}, {\bf 3}(1):69-72, 2013.

\bibitem{redner_book} Pavel L. Krapivsky and Sidney Redner and Eli
  Ben-Naim \newblock {\it A Kinetic View of Statistical Physics}.
  \newblock {Cambridge University Press, Cambridge, UK}, 2010.


\bibitem{Ziff 1980}
Ziff R. M. and Stell G.
\newblock {\it Kinetics of Polymer Gelation}.
\newblock {The Journal of Chemical Physics}, {\bf 73}(7):3792, 1980.


\bibitem{wattis 1998}
Wattis Jonathan A. D. and King John R.
\newblock {\it Asymptotic solutions of the Becker-Döring}.
\newblock {Journal of Physics A: Mathematical General}, {\bf 31}:7169-9189, 1998.

\bibitem{Engelmann 2006}
Bernd Englemann
\newblock {\it Initiation of coagulation by tissue factor carriers in blood}.
\newblock {Blood Cells, Molecules, and Diseases }, {\bf 36}:188-190, 2006.

\bibitem{Hockin 2002}
Matthew F. Hockin, Keneth C. Jones, Stephen J. Everse, Kenneth G. Mann
\newblock {\it A model for the stoichiometric regulation of blood coagulation}.
\newblock {The Journal of Biological Chemistry}, {\bf 227}(21):18322-18333, 2002.

\bibitem{Chatterjee 2010}
Manash S. Chatterjee, William S. Denney, Huiyan Jing, Scott L. Diamond
\newblock {\it System biology of coagulation initialtion: Kinetics of thrombin generation in resting and activated human blood}.
\newblock {PLOS Computational Biology}, {\bf 6}(9):e1000950, 2010.

\bibitem{Guy 2005}
Robert D. Guy, Aaron L. Fogelson, James P. Keener
\newblock {\it Fibrin gel formation in a shear flow}.
\newblock {Mathematical Medicine and Biology}, {\bf 0}:1-20, 2005.


\bibitem{Mounts 1997}
William M. Mounts, Michael N. Liebman
\newblock {\it Qualitative modeling of normal blood coagulation and its pathological states using stochastic activity networks}.
\newblock {International Journal of Biological Macromolecules}, {\bf 20}:265-281, 1997.



\bibitem{hoze 2016} Hoze Nathanael and Holcman David \newblock {\it
  Stochastic coagulation-fragmentation processes with a finite number
  of particles and applications}.  \newblock {Annals of Applied
  Probability}, {\bf 28}(3):1449-1490, 2016.


\bibitem{niethammer 2003}
Niethammer B.
\newblock {\it On the Evolution of Large Clusters in the Becker-Döring Model}.
\newblock {Journal of Nonlinear Science}, {\bf 13}(1):115-155, 2003.


\bibitem{penrose 1997} Penrose O..  \newblock {\it The Becker-Döring
  equations at large times and their connection with the LSW theory of
  coarsening}.  \newblock {Journal of Statistical Physics}, {\bf
  89}(1-2):305-320, 1997.


\bibitem{davis 2016}
Davis J. K. and Sindi S. S.
\newblock {\it Initial condition of stochastic self-assembly}.
\newblock {Physical Review E}, {\bf 93}(2):022109, 2016.


\bibitem{Gupta 2017} Gupta Ankit, Mikelson Jan and Khammash Mustafa
  \newblock {\it A Finite State Projection Algorithm for the
    Stationary Solution of the Chemical Master Equation}.  \newblock
            {The Journal of Chemical Physics}, {\bf 147}:154101,
            2017.

\bibitem{Sudbrack 2015} Vítor Sudbrack, Leonardo G.Brunnet, Rita M.C
  de Almeida, Ricardo M. Ferreira, Daniel Gamermann \newblock {\it
    Master Equation for Degree Distribution of a Duplication and
    Divergence Network}.  \newblock {Physica A}, {\bf 509}:588-298,
  2015.


\bibitem{Smadbeck 2014} Smadbeck Patrick and Kaznessis Tiannis N.
  \newblock {\it Solution of Chemical Master Equations for Nonlinear
    Stochastic Reaction Networks}.  \newblock {Current Opinion in
    Chemical Engineering.}, {\bf 5}:90-95, 2014.


\bibitem{cao 2013} Cao Youfang and Liang Jie \newblock {\it Adaptively
  Biased Sequential Importance Sampling for Rare Events in Reaction
  Networks with Comparison to Exact Solution from Finit Buffer dCME
  Method}.  \newblock {The Journal of Chemical Physics}, {\bf
  139}(2):025101, 2013.



\bibitem{Terebus 2014} Terebus Anna, Cao Youfang and Liang Jie
  \newblock {\it Exact Computation of Probability LAndscape of
    Stochastic Networks of Single Input and Coupled Toggle Switch
    Modules}.  \newblock {IEEE Engineering in Medicine and Biology
    Society (EMBS)}, {\bf 2014}5228-5231, 2014.


\bibitem{cao 2016A} Cao Youfang, Terebus Anna and Liang Jie \newblock
  {\it State Space Truncation with Quantified Errors for Accurate
    Solution to Discrete Chemical Master Equation}.  \newblock
  {Bulletin of Mathematical Biology}, {\bf 78}(4): 617-661, 2016.


\bibitem{D'orsogna 2012}
D’Orsogna Maria R., Lakatos G. and Chou Tom
\newblock {\it Stochastic Self-Assembly of Incommensurate Clusters}.
\newblock {The Journal of Chemical Physics}, {\bf 136}:0884110, 2012.



\bibitem{D'orsogna 2015}
D’Orsogna Maria R., Lei Qi and Chou Tom
\newblock {\it First Assembly Times and Equilibrium in Stochastic Coagulation-Fragmentation}.
\newblock {The Journal of Chemical Physics}, {\bf 143}:014112, 2015.


\bibitem{kotalczyk 2017} Kotalczyk G. and Kruis F. E.  \newblock {\it
  A Monte Carlo Method for the Simulation of Coagulation and
  Nucleation Based on Weighted Particles and the Concepts of
  Stochastic Resolution and Merging}.  \newblock {Journal of
  Computational Physics}, {\bf 340}:276-296, 2017.


\bibitem{smith 2018}
Smith Alastair J., Wells Clive G. and Kraft Markus
\newblock {\it A New Iterative Scheme for Solving the Discrete Smoluchowski Equation}.
\newblock {Journal of Computational Physics}, {\bf 352}:373-387, 2018.

\bibitem{cao 2016B}
Cao Youfang, Terebus Anna and Liang Jie
\newblock {\it Accurate Chemical Master Equation Solution Using Multi- Finite Buffers}.
\newblock {SIAM Multiscale Modeling and Simulation}, {\bf 14}(2): 923-963, 2016.


\bibitem{cao 2008} Cao Youfang and Liang Jie \newblock {\it Optimal
  Enumeration of State Space of Finitely Buffered Stochastic Molecular
  Networks and Exact Computation of Steady State Landscape
  Probability}.  \newblock {BMC Systems Biology}, {\bf 2}(30):1-13,
  2008.

\bibitem{cao 2016C} Cao Youfang, Terebus Anna and Liang Jie \newblock
  {\it Chapter 3. Modeling Dtochastic Gene Regulatory Networks Using
    Direct Solutions of Chemical Master Equation and Rare Event
    Sampling, Book title: Research in Analysis and Modeling of Gene
    Regulatory Networks, Ivanov Ivan V., Qian Xiaoning and Pal
    Ranadip}.  \newblock {Advance in Medical Technologies and Clinical
    Practice (AMTCP)}, {\bf Book Series 2016}.


\bibitem{FPT_REVIEW} Tom Chou and Maria R. D'Orsogna \newblock {\it
  First passage Problems in Biology} in \newblock {Chapter 13:
  First-Passage Phenomena and Their Applications, editors Ralf
  Metzler, Gleb Oshanin, and Sidney Redner}, \newblock {World
  Scientific, 2014}

\bibitem{YVINEC} R. Yvinec and Maria R. D'Orsogna and Tom Chou.
\newblock {\it First passage times in homogeneous nucleation and
  self-assembly}.  
\newblock {The Journal of Chemical Physics}, {\bf 137}: 244107, 2012.

\end{thebibliography}
\end{document}